\documentclass{PoS}
\title{Recent progress in Hamiltonian light-front QCD}

\ShortTitle{Recent progress in Hamiltonian light-front QCD}

\author{\speaker{J.~P.~Vary}
         \\
        Iowa State University, Ames, Iowa, USA\\
        E-mail: \email{jvary@iastate.edu}}

\author{H. Honkanen, Jun Li, P.~Maris\\
Iowa State University, Ames, Iowa, USA}

\author{S.~J.~Brodsky\\
SLAC National Accelerator Laboratory, Stanford University, Menlo Park, California, USA}

\author{P.~Sternberg, E.~G.~Ng, C.~Yang\\
Lawrence Berkeley National Laboratory, Berkeley, California, USA}

\abstract{Hamiltonian light-front quantum field theory 
constitutes a framework for the non-perturbative solution of invariant masses and correlated 
parton amplitudes of self-bound systems. By choosing light-front gauge and adopting
a basis function representation, we obtain a large, sparse, Hamiltonian matrix
for mass eigenstates of gauge theories that is solvable by adapting the {\it ab initio} no-core methods 
of nuclear many-body theory. Full covariance is recovered in the continuum limit, the infinite matrix limit. We outline our approach and discuss the computational challenges.}

\FullConference{LIGHT CONE 2008 Relativistic Nuclear and Particle Physics\\
                 July 7-11, 2008\\
                 Mulhouse, France}

\begin{document}

\section{Introduction}
Non-perturbative Hamiltonian light-front quantum field theory presents opportunities 
and challenges that bridge particle physics and nuclear physics.
Major goals include predicting both the masses and transitions rates of the
hadrons and their structures as seen in high-momentum transfer experiments.
Current focii of intense experimental and theoretical research that could 
benefit from insights derived within this Hamiltonian approach 
include the spin structure of the proton, the neutron electromagnetic form factor,
the generalized parton distributions of the baryons, etc.

Hamiltonian light-front field theory 
in a discretized momentum basis \cite{BPP98} and in transverse lattice
approaches \cite{JPV214,GP08} have shown significant promise.
We present here a basis-function approach that exploits
recent advances in solving the non-relativistic strongly interacting nuclear
many-body problem \cite{NCSM, NCFC}.  We note that both fields of physics
face common issues within the Hamiltonian approach - i.e. how to 
(1) define the Hamiltonian;
(2) renormalize to a finite space;
(3) solve for non-perturbative observables while preserving all symmetries; and,
(4) take the continuum limit.

We begin with a brief overview of recent advances in solving 
light nuclei with realistic nucleon-nucleon (NN) and three-nucleon (NNN)
interactions using {\it ab initio} no-core methods. 
Then, we introduce our basis function approach to
light-front QCD within the light-front gauge.  
Renormalization/regularization issues are also addressed.
We present illustrative features of our approach with the example
of cavity-mode QED and sketch its extension to cavity-mode QCD.

\section{No Core Shell Model (NCSM) and No Core Full Configuration (NCFC) methods}

To solve for the properties of nuclei, self-bound strongly interacting systems, with
realistic Hamiltonians, one faces immense analytical and computational
challenges.  Recently, {\it ab initio} approaches have been developed that
preserve all the underlying symmetries and they 
converge to the exact result.  The basis function approach that we adopt here
\cite{NCSM, NCFC} is one of several methods shown to be successful.  
The primary advantages are its flexibility for choosing the Hamiltonian, the method
of renormalization/regularization and the basis space.  
These advantages impel us to adopt the basis function approach 
in light-front quantum field theory.

While non-relativistic applications in finite nuclei restrict the basis to a fixed number
of fermions, we introduce here the extension to a flexible number of fermions,
antifermions and bosons.

Refs. \cite{NCSM} and \cite{NCFC} provide examples of the recent advances in the {\it ab initio} NCSM and NCFC, respectively.  The former adopts 
a basis-space renormalization method and applies it to realistic NN and NNN interactions (derived from chiral effective field theory) to solve nuclei with Atomic Numbers 10-13 \cite{chiral07}. Experimental binding energies, spectra, electromagnetic moments and transition rates are well-reproduced.
The latter adopts a realistic NN interaction that is sufficiently soft  that binding energies from a sequence of finite matrix solutions may be extrapolated to the infinite matrix limit.  Again, one obtains good agreement with experiment.

It is important to note the analytical and technical advances made to solve these problems. First, non-perturbative renormalization has been developed to accompany these basis-space methods and their success is impressive. Several schemes have emerged and current research focuses on
detailed understanding of the scheme-dependence of convergence rates (different observables converge at different rates) \cite{Bogner07}. Second, large scale calculations are performed on leadership-class parallel computers to solve for the low-lying eigenstates and eigenvectors as well as to carry out evaluation of a suite of experimental observables.  Low-lying solutions for matrices of dimension 2-billion on 30,628 processors with a 3-hour run is the current record.  However, one expects these limits to be exceeded very soon as the techniques are evolving rapidly \cite{Sternberg08} and the computers are growing dramatically.  Matrices with dimensions in the tens of billions with strong interaction Hamiltonians will soon be solvable.

In a NCSM or NCFC application, one adopts a 3-D harmonic oscillator for all the particles in the nucleus (with harmonic oscillator energy $\hbar\Omega$), treats the neutrons and protons independently, and generates a many-fermion basis space that includes the lowest oscillator configurations as well as all those generated by allowing up to $N_{max}$ oscillator quanta of excitations.  The single particle states specify the orbital angular momentum projection and the basis is referred to as the $m$-scheme basis. For the NCSM one also selects a renormalization scheme linked to the basis truncation while in the NCFC the renormalization is either absent or of a type that retains the infinite matrix problem.  In the NCFC case \cite{NCFC}, one extrapolates to the continuum limit as we now illustrate.

\begin{figure}[!ht]
\includegraphics[width=11cm]{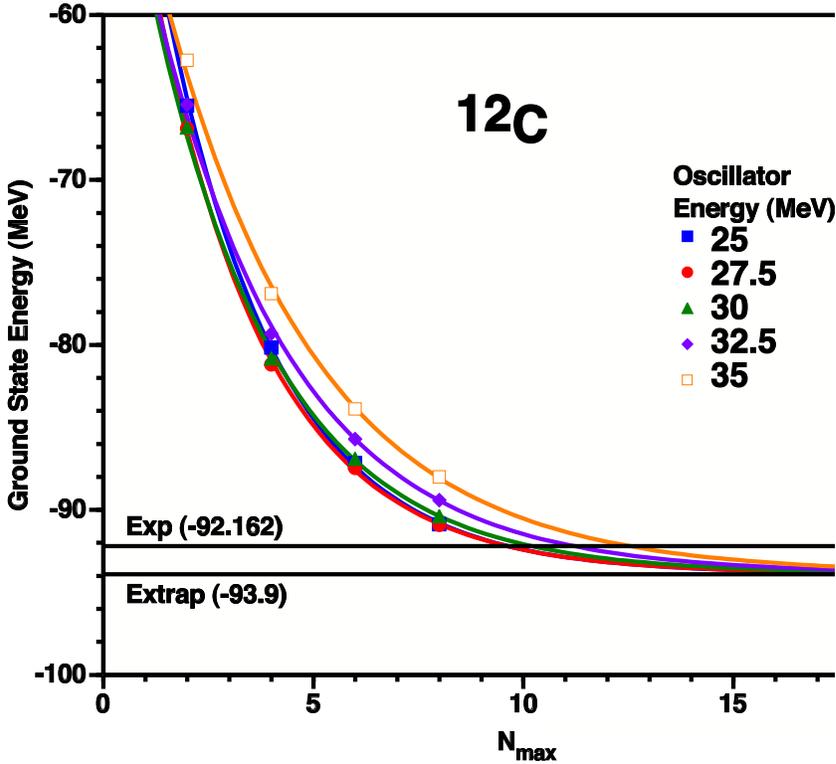}
\caption{Calculated ground state energy of $^{12}$C for
$N_{max}=2{-}8$ (discrete points) at selected values of $\hbar\Omega$.  
For each $\hbar\Omega$, the results are fit to an exponential plus
a constant, the asymptote, which is constrained to be the same for 
each curve\cite{NCFC}.  We display the experimental
ground state energy and the common asymptote.} 
\label{12C}
\end{figure}

We show in fig. \ref{12C} results for the ground state of $^{12}C$ as a function of $N_{max}$
obtained with a realistic NN interaction, JISP16 \cite{Shirokov07}.  The smooth curves portray fits that achieve the desired independence of  $N_{max}$ and $\hbar\Omega$ so as to yield the extrapolated ground state energy.   Our assessed uncertainty in the extrapolant is about 2 MeV so there is rather good agreement with experiment.  The largest cases correspond to $N_{max}=8$, where the matrix reaches a dimension near 600 million.  $N_{max}=10$ produces a matrix near 8 billion which we plan to solve in the coming year.

\section{Cavity mode light-front QED and QCD}

It has long been known that light-front Hamiltonian quantum field theory has similarities with non-relativistic quantum many-body theory.  We further exploit this connection, in what we will term a "Basis Light Front Quantized (BLFQ)" approach, by adopting a light-front basis space consisting of the 2-D harmonic oscillator for the transverse modes (radial coordinate $\rho$ and polar angle $\phi$) and a discretized momentum space basis for the longitudinal modes with anti-periodic boundary conditions (APBC).  The 2-D oscillator states are characterized by their principal quantum number $n$ and orbital quantum number $m$.   Adoption of this basis is also consistent with recent developments in AdS/CFT correspondence with QCD \cite{Brodsky}.



\begin{figure}[!ht]
\includegraphics[width=8cm,angle=90]{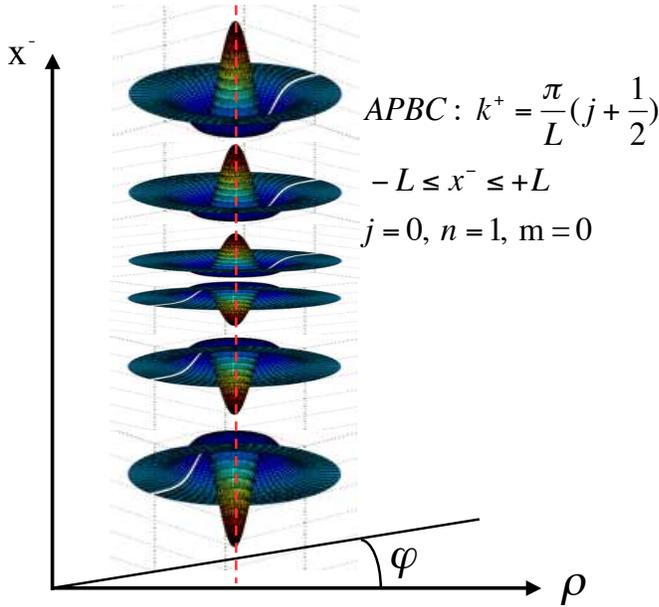}
\caption{Transverse sections of a 3-D basis function involving a 2-D harmonic oscillator and a longitudinal mode with antiperiodic boundary conditions (APBC).} 
\label{3Dview}
\end{figure}

In order to bring the full 3-D basis into perspective, we select a transverse and the lowest longitudinal mode (with index $j=0$) and display slices of the 3-D basis function at selected longitudinal coordinates, $x^-$ in fig. \ref{3Dview}.  Our purpose in presenting figs. \ref{3Dview} is to suggest the richness and economy of texture available for solutions in a basis  function approach.  Note that the choice of basis functions is arbitrary except for the standard conditions of orthonormality and completeness.

\begin{figure}[!ht]
\includegraphics[width=11cm]{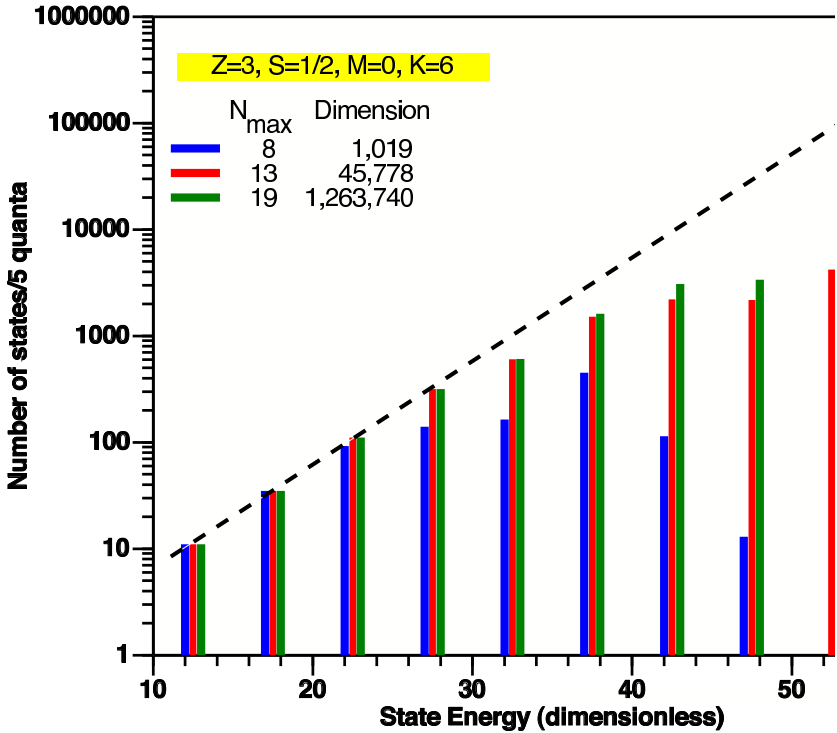}
\caption{State density from BLFQ for 3 identical massless fermions confined in a trap for
a selection of $N_{max}$ values at fixed $K=6$.  The dimensions of the
resulting matrices are presented in the legend.  The states are binned in groups of 
5 units of energy (quanta) where each parton carries quanta equal to its 2-D oscillator energy divided by its light-front momentum fraction ($x_i=(j_i+1)/K$). The dashed straight line traces the exponential increase in state density, familiar in many-body theory, when the basis is sufficiently large.} 
\label{histogram}
\end{figure}

To further illustrate the BLFQ approach, we consider a non-interacting QED system confined to a transverse harmonic trap or cavity.   For simplicity, we take the leptons as massless. The basis functions are matched to the trap so we implement a transverse 2-D harmonic oscillator basis with length scale fixed by the trap and an infinite system in the longitudinal direction with APBC.  The symmetries (and their chosen values for the case illustrated in fig. \ref{histogram}) imposed on the many-parton basis states are total charge (Z=3), total magnetic projection ($M=0$) in the transverse space, total spin projection ($S=1/2$), maximum total 2-D oscillator quanta ($N_{max}=8-19$) and total longitudinal momentum in dimensionless units ($\sum{j_i+1}=K=6$).  The range of the number fermion-antifermion pairs and bosons is limited by the cutoffs in the modes ($N_{max}$ and $K$): e.g. each parton carries at least one unit of longitudinal momentum so the basis  consists of up to 6 partons.

Fig. \ref{histogram} demonstrates the saturation of low-lying unperturbed modes with increasing $N_{max}$.  Already, this state density could serve as input to a model for the statistical mechanics of the system treated in the microcanonical ensemble.  Of course, interactions must be added to make the model realistic at low temperatures where correlations are important.  After turning on the interactions, the challenge will be to evaluate observables and demonstrate convergence with respect to the cutoffs ($N_{max}$ and $K$).  Independence of the basis scale, $\hbar\Omega$, must also be obtained.  These are the standard challenges of taking the continuum limit.

The Hamiltonian $H$ for this system ($KH$ gives the invariant mass-squared) is defined by the sum of the occupied modes with the scale set by the combined constant $\Lambda^2 = 2M_0\hbar\Omega$:
$$
H = 2M_0 P^-_c = \frac{2M_0\hbar\Omega}{K}\sum{\frac{2n_i+|m_i| +1}{x_i}}
$$

\begin{figure}[!hl]
\includegraphics[width=11cm,angle=90]{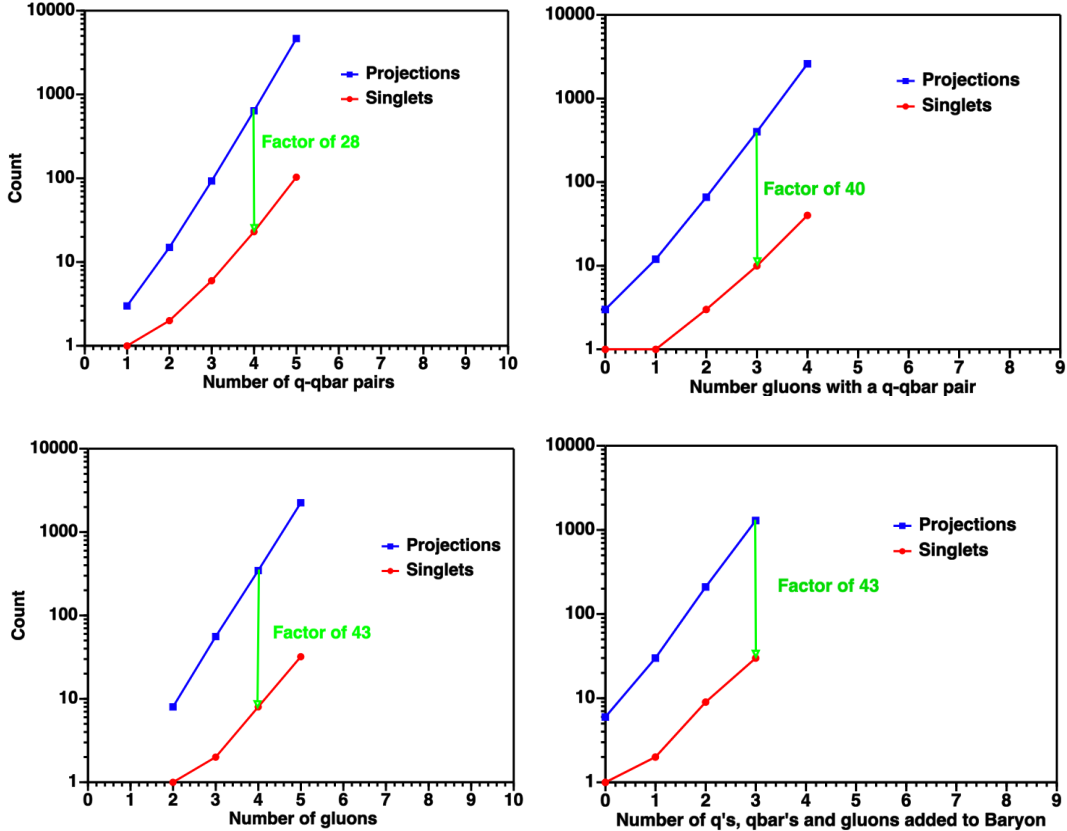}
\caption{Number of color space states that apply to each space-spin configuration of selected multi-parton states for two methods of enumerating the color basis states. The upper curves are counts of all color configurations with zero color projection.  The lower curves are counts of global color singlets.} 
\label{colorstates}
\end{figure}

We can extend the approach to QCD by implementing the SU(3) color degree of freedom for each parton - 3 colors for each fermion and 8 for each boson.   We consider two versions of implementing the global color singlet constraint.   In the first case, we follow Ref. \cite{Lloyd} by constraining all color components to have zero color projection and adding a Lagrange multiplier term to the Hamiltonian to select global color singlet eigenstates.  This results in the upper curves in fig. \ref{colorstates}.  In the second case, we restrict the basis space to global color singlets and this results in the lower curves in fig. \ref{colorstates}.  The second method produces a factor of 30-40 lower multiplicity at the upper ends of these curves at the cost of increased computation time for matrix elements. Either implementation dramatically increases the state density over the case of QED, but the use of a global color singlet constraint is clearly more effective in minimizing the explosion in basis space states.

Now we briefly address the interacting theory. We have identified the appearance of the expected log divergence in the matrix element of the fermion-boson vertex as it falls only as the inverse square root of the principal quantum number in selected sequences.  We are investigating sector-dependent coupling constant renormalization to manage this divergence \cite{Karmanov}.

\section{Conclusion
\label{concl}}
Following successful methods of {\it ab initio} nuclear many-body theory, we have introduced a basis light-front quantization (BLFQ) approach to Hamiltonian quantum field theory and illustrated some of its key features with a cavity mode treatment of QED.  We indicated our method of treating color for extending our approach to QCD.  The computational requirements of this approach are substantial, and we foresee extensive use of leadership-class computers to obtain practical results.

\acknowledgments The authors thank A. Harindranath, V. 
Karmanov and J. Hiller for fruitful discussions.  This work was supported in part 
by the U.S. Department of Energy Grant DE-FG02-87ER40371.

\end{document}